\documentclass[twocolumn,amsmath,amssymb,pre,floatfix]{revtex4}

\usepackage{color}
\usepackage{graphicx}
\usepackage{dcolumn}
\usepackage{times}
\usepackage{float}

\begin{document}

\title{Multiplicity of actuated shapes in woven fabrics with twisted Janus fibres}
\author{A.~P. Zakharov, L.~M. Pismen}
\affiliation{Department of Chemical Engineering, Technion -- Israel Institute of Technology, Haifa 32000, Israel}

\begin{abstract}
We investigate actuation of woven fabrics including active Janus fibres with an imposed twist, which bend in variable directions upon phase transition between isotropic and nematic state. The essential feature of textiles incorporating a pair of Janus fibres with a mismatched pitch or handedness of coiling is the existence of multiple stable shapes with different energies within a certain range of the extension coefficient. If the active fibres are closed into a ring, torsion develops to accommodate adjustment of the direction of bending. The structure is generally stabilised by adding more passive filaments, and multistability is observed also in this case.
\end{abstract}
 
\maketitle
 
\section{Introduction}

In an earlier publication \cite{textile}, we described a variety of shapes emerging in textiles containing \emph{Janus fibres} which bend when their active component changes its length upon actuation causing the phase transition to nematic state oriented along its axis. The natural direction of bending is perpendicular to the dividing plane between the active and passive components of the fibre, which was assumed in the earlier study to be constant. More interesting shapes can be obtained if  the Janus fibre is \emph{twisted}, so that the direction of bending changes along its length. Deformations of \emph{single} naturally curved twisted filaments (birods) have been studied in a number of publications \cite{GorielyTabor,GorielyGoldstein,Liu15,Audoly17,Goriely17}. Instabilities of closed twisted loops have a long history reviwed by Goriely \cite{Goriely06}, starting from the early work by Michell \cite{Michell}. In our recent publication \cite{Jring}, we have shown that alternative twisted configurations, some of them metastable, can be obtained by perturbing a ring in different ways, and compared their energies.  

\begin{figure}[t]
\centering
 \includegraphics[width=.25\textwidth]{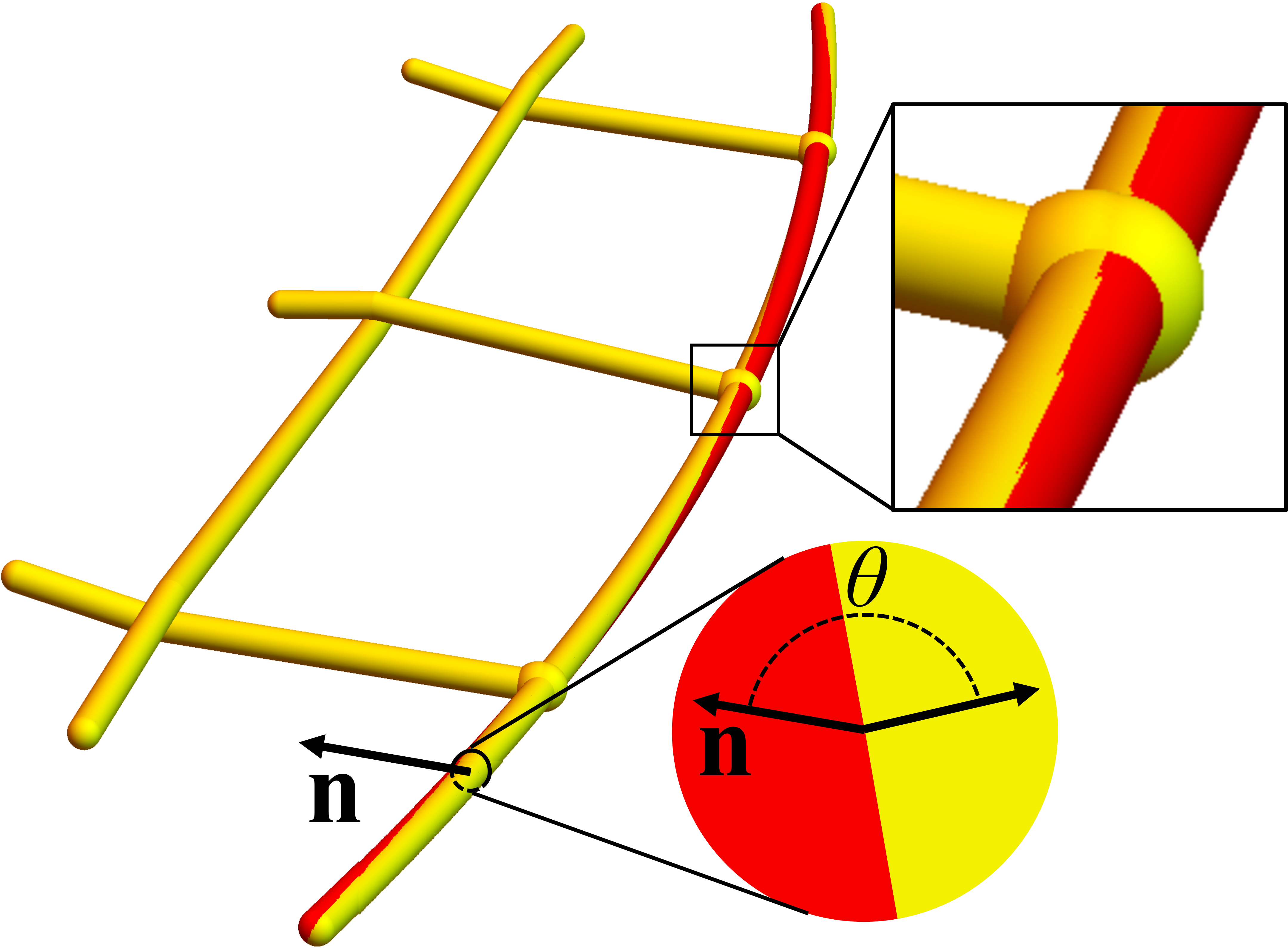}\\
\caption{A piece of the textile structure with noose connections of passive filaments to a Janus fibre and the cross-section of a Janus fibre with a mismatched orientation of the normal vector \textbf{n} to the centerline and the internal normal.}  
\label{sketch} 
\end{figure}

This communication aims at merging both directions of research by incorporating twisted Janus fibres into woven fabrics of passive filaments. Twisted Janus filaments can be fabricated using two connected extruders supplying nematic and isotropic elastomers similar to untwisted ones but adding simultaneous melt spinning \cite{IonovJanus}. Unlike textiles with untwisted active fibres, rigidly attaching passive fibres to active ones would cause them to strongly curve at these junctions. Instead, we assume noose connections allowing for free rotation but not gliding along active filaments (Fig.~\ref{sketch}). The passive fibres are allowed, as in the earlier study~\cite{textile}, to mutually slide at their intersections in the woven fabric. We will further compute energies of different stable configurations and investigate transitions between stable and metastable states. We will see, in particular, that a woven network stabilises configurations of closed loops, which are absolutely unstable when isolated \cite{Jring}.       

\section{Basic equations} 

We consider a woven fabric combining passive elastic filaments and twisted Janus fibres of the same radius $r$. The latter consist of an active component occupying the sector $|\phi|>\pi/2$ of their circular cross-section and a passive component in the sector $|\phi|<\pi/2$.  (Fig.~\ref{sketch}). The active component  is assumed to be a nematic elastomer that polarises upon phase transition along the filament axis thereby causing elongation by a factor $\lambda=1+\epsilon$, while the passive component in the other half-section remains unchanged.  
 
When the nematic component elongates, the centreline of the filament develops an intrinsic curvature, which, when unforced and unconstrained, is directed along the normal to the dividing plane. The curvature radius $R$ is proportional to $r/\epsilon$, and therefore small extensions are sufficient to strongly bend a thin filament in a direction rotating due to imposed twist \cite{textile}. The direction of bending may change in the presence of constraints, in particular, when the filament forms of a closed loop or its bending is restricted by ensuing deformation of attached fibres forming the textile network.

We define the elastic energy of an inextensible filament in the standard slender body (Kirchhoff rod) approximation \cite{LL,AP} as the integral along its centerline (parametrised by the arc length $s$) that includes both flexural and torsional energies, respectively, $F^b$ and $F^t$:  
\begin{equation}
\mathcal{F}^e =  \frac12 E A \int ( F^b + F^t )ds , 
\label{Fe}
\end{equation}
where $A$ is the cross-sectional area and $E$ is the Young modulus. The flexural rigidity of active filaments is computed \cite{Jring} by adding up the contributions of the active ($I_a$) and passive ($I_p$) sectors:
\begin{align}
&I_a =\frac{1}{\pi r^2}\int_0^r r \,d r 
\int_{\pi/2}^{3\pi/2} \left(\frac rR \cos (\phi-\theta) -\epsilon\right)^2 d \phi \notag \\
 &= \frac 18  \left(\frac{r}{R}\right)^2 - \frac{4\epsilon}{3\pi}\frac{r}{R}\cos\theta +  \frac{\epsilon^2}{2} ,
  \label{Iin} \\
&I_p =\frac{1}{\pi r^2}\int_0^r \frac {r^3}{R^2} \,d r  
\int^{-\pi/2}_{\pi/2}\cos^2 (\phi-\theta) \,d \phi
= \frac 18  \left(\frac{r}{R}\right)^2, \label{Iout} \\
&F^b (\kappa,\theta) = I_a+I_p = \frac 14  \left(\kappa r\right)^2 
 -\frac {4\,\epsilon}{3 \pi} \kappa  r \cos\theta+\frac {\epsilon^2}{2} .
 \label{Ipi}
\end{align}
where $\kappa=1/R$ is curvature. This expression depends on the angle $\theta$ between the normal to the dividing plane (to be called internal normal) and the curvature vector \textbf{n} (see Fig.~\ref{sketch}). The lowest energy state is, clearly, attained at $\theta=0$ when the directions of both vectors coincide. The natural (intrinsic) curvature upon actuation $\widehat{\kappa}$ is determined by the condition of minimum overall strain energy in the cross-section $dF^b/d\kappa=0$, yielding $\widehat{\kappa} r= \frac 83 \epsilon/\pi  \approx 0.849 \epsilon$. The residual energy at the optimal bending is of the order $O(\epsilon^2)$. This formula with modified numerical coefficients and appropriate values of  $\widehat{\kappa}$ is applicable also to filaments with different cross-sectional shapes and distributions of the active and passive components. The flexural rigidity of a passive filament is defined by the standard expression \cite{LL,AP} $I_p= \frac {1}{4}(r/R)^2 $.  

In the absence of rigid connections between filaments, there are no externally imposed torques but an extra twist $\tau$ modifying that imposed in the manufacturing process may emerge spontaneously in closed active filaments (Janus rings \cite{Jring}) as they tend to align the direction of bending with the internal normal (more on this in Sect.~\ref{ring}). The torsional rigidity is defined by the standard expression $F^t=\frac 12 (r \tau)^2$.  Since filaments are not firmly attached one to the other, mutually imposed torques are absent and no torsion arises in open-ended filaments. 

Equilibrium configurations minimise the total energy of the system, and can be attained following the pseudo-time evolution equations for the positions of the centerlines of each filament $\mathbf{x}_i(s)$ and the local orientation angles $\theta_i(s)$:
 \begin{equation}
\frac{d \mathbf{x}_i(s)}{d t}= - \frac{\delta\mathcal{F}}{\delta \mathbf{x}_i(s)},
  \qquad \frac{d \theta(s)}{d t}= - \frac{\delta\mathcal{F}}{\delta \theta(s)}.
 \label{evol}
\end{equation}
These equations are discretized as 
 \begin{align}
\frac{d \mathbf{x}_{ij}}{d t}&= - \frac{\partial}{\partial \mathbf{x}_{ij}} 
 \sum_{i,j}({F}^{s}_{ij}+{F}^{b}_{ij} ), \notag \\
\frac{d \theta_{ij}}{d t}&= - \frac{\partial}{\partial \theta_{ij}}   
 \sum_{i,j}({F}^{b}_{ij}+{F}^{t}_{ij}) .
 \label{evold}
\end{align}
The evolution equation of $\theta$ in Eqs.~\eqref{evol}, \eqref{evold} is implemented only for closed active fibres, otherwise $\theta$ is retained at its initially imposed values changing along a twisted filament. For passive filaments this equation is never relevant. The reduced axial strain energy per discretised segment 
 \begin{equation}
{F}^s_{ij} = \frac{|\mathbf{x}_{ij}-\mathbf{x}_{i,j-1}|^2}{\delta s^2} ,
\label{strain}
\end{equation}
where $\delta s$ is the discretisation length, is added here to suppress the axial extension in numerical computations to levels not exceeding $\kappa r \ll1$ by the order of magnitude. Since extensions upon phase transition have to be small, slight variations of the cross-sectional area can be neglected. The positions $\mathbf{x}_{ij}$ are constrained by connections at the textile nodes. Self-intersections at originally far removed locations are checked and prevented in the course of simulations. The woven structure of the textile imposes a ``microcurvature" component along the normal to the envelope surface of the fabric at intersection nodes, that depends on the distance between intersections and prevents convergence of neighbouring nodes to distances comparable to the diameter of the fibres. This introduces additional bending energy, and therefore the energy of textiles in the following computations does not vanish at $\epsilon=0$. The associated length changes can be neglected. 
\section{Shape multistability of a strip \label{strip}}

We start with reshaping of a rectangular textile strip embroidered over the long sides by two active naturally twisted fibres, with passive filaments forming the rest of the connecting fabric; all fibres have the same radius and mechanical properties. The twisted fibres may differ by pitch and coiling handedness. Here we focus on some basic cases, excluding variations of filament thickness, but putting the emphasis on the emergence of multiple solutions. 
	
The case of a textile strip is framed by two twisted Janus filaments of the same pitch and handedness is trivial. This configuration is rather exceptional, as it does not break the symmetry and therefore cannot lead to variegated shapes. Since the Janus fibres can easily rotate about their axes to reduce the bending energy, the active filaments acquire the same orientation when they bend upon actuation in the direction aligned with the local internal normal, which rotates along the filament with the imposed pitch. No torsion arises in this configuration, as in all cases when neither closed filaments nor rigid connections are present. The mutual distance between the active fibres slightly changes due to the resistance of parallel passive filaments, which deform to adjust to the helical shape of the active fibres and thereby force perpendicular filaments.  

	\begin{figure}[t]
		\begin{tabular}{cc} 
		(a)&(b)\\
		\includegraphics[width=.225\textwidth]{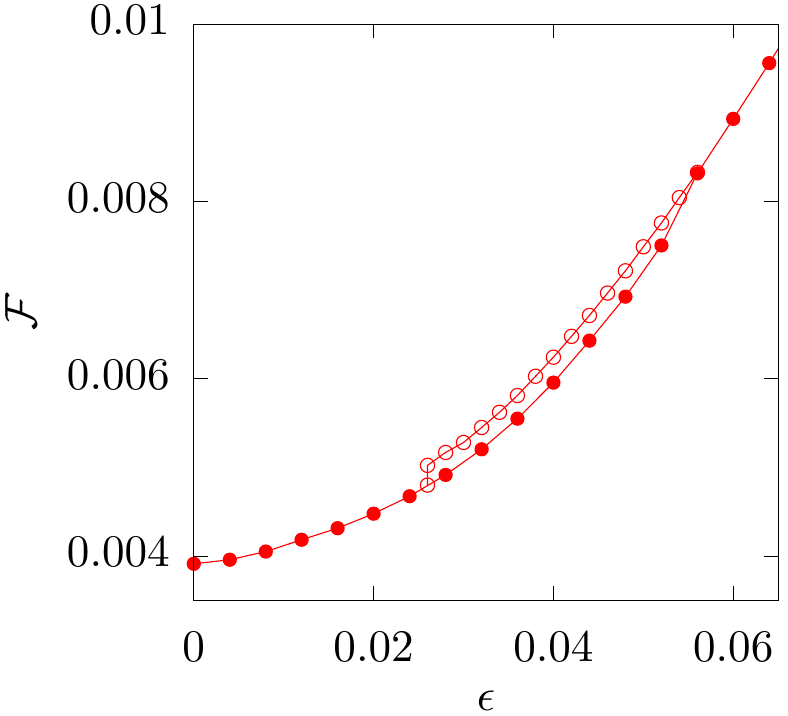}
		&\includegraphics[width=.225\textwidth]{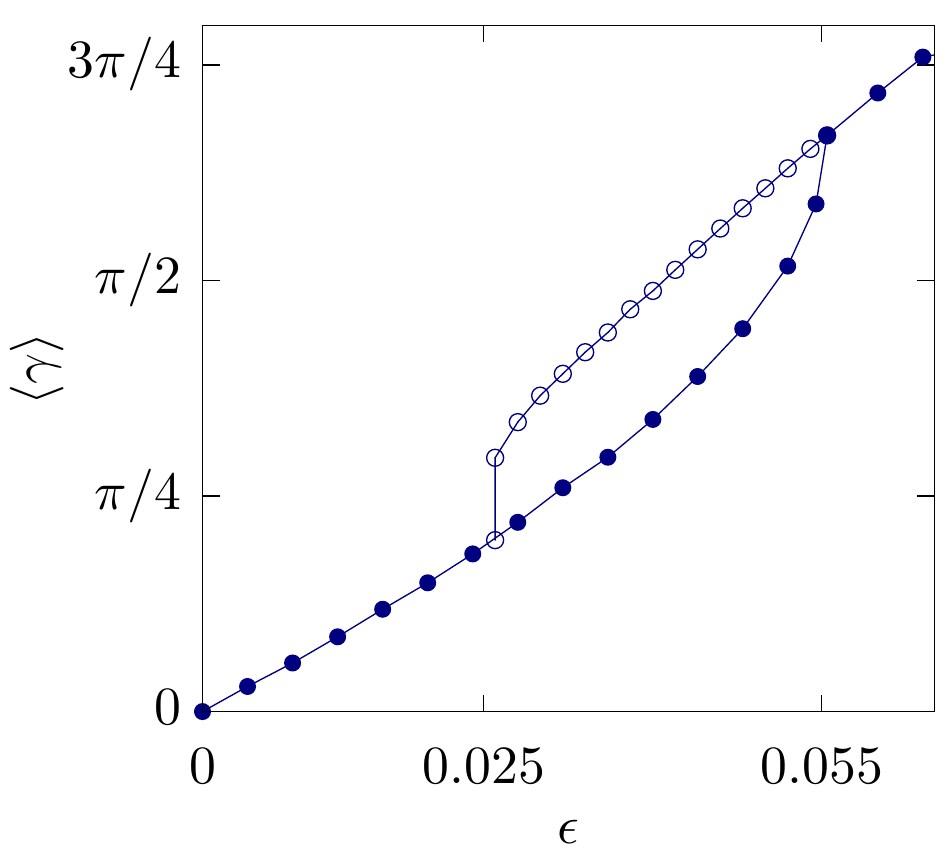}\\
		(c)&(d)\\
		\includegraphics[width=.225\textwidth]{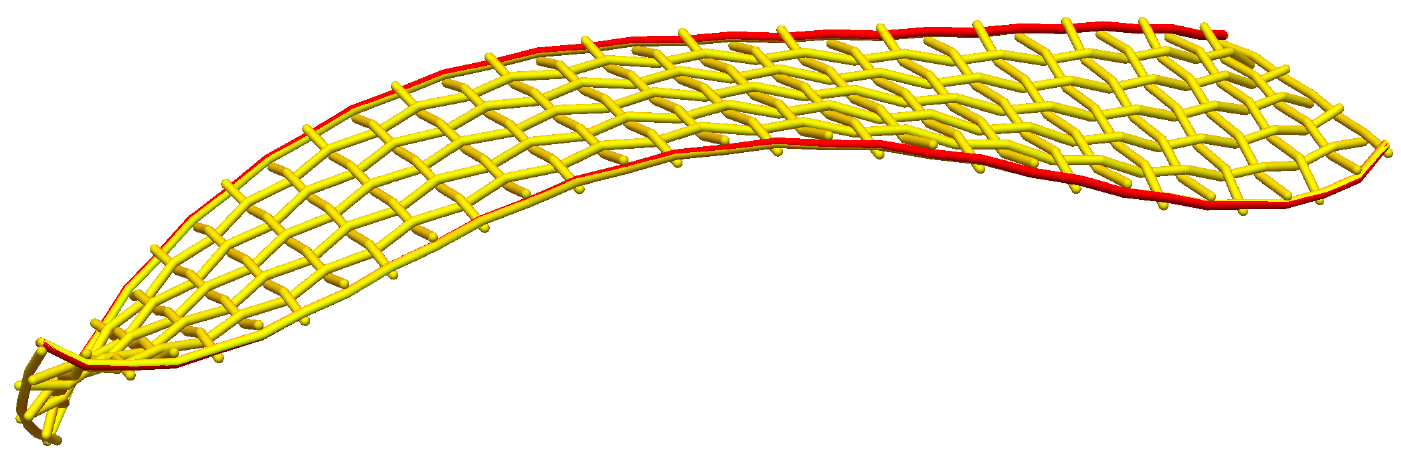}
		&\includegraphics[width=.225\textwidth]{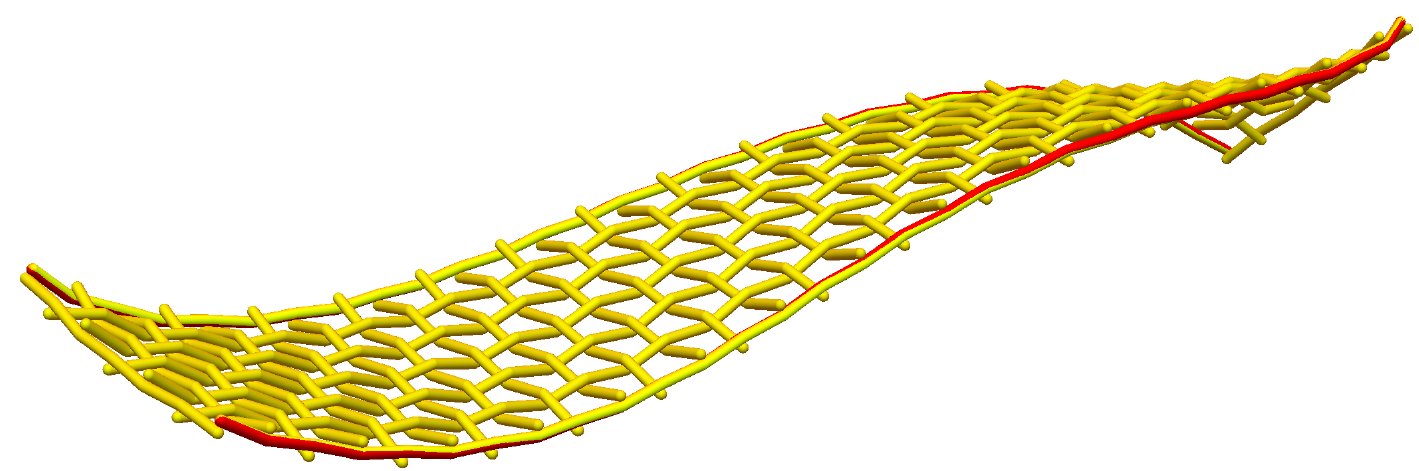}
		\end{tabular}
		\caption{The overall energy (a) and the average deviation angle $\langle \gamma \rangle$ (b) of two active fibres with the same handedness and different pitch. The equilibrium shapes of the stable (c) and metastable (d) configurations are shown at $\epsilon=0.03$.}
		\label{StripSame}
	\end{figure}             
	 
The symmetry is broken when Janus fibres have different pitch. This case is more interesting, as their intrinsic curvature vectors become misaligned upon actuation and, as a result, the distances between initially parallel active fibres become mismatched.  We found that two fibres with the total rotation of the internal normal $\pi$ and $2\pi$ cause the textile to modify its shape in a qualitatively significant way as the parameter $\epsilon$ is varied. Figs.~\ref{StripSame}a,b show the change of the overall energy $\mathcal{F}$ and of the overall shape of the textile with $\epsilon$. The distinction is better seen in Figs.~\ref{StripSame}b. There are many ways of choosing a single parameter to characterise the overall shape. Our choice is the angle $\langle \gamma \rangle$ between the average of the directions of the internal normals of the two Janus fibres at both ends and the average direction of all longitudinal fibres of the textile ribbon. An example of distinct equilibrium shapes at the same value of $\epsilon$ is shown in Fig.~\ref{StripSame}c,d. Although the changes in the overall energy are small, the shapes differ significantly. In a wide interval of $\epsilon$, there exist two alternative states, one of them absolutely stable (Fig.~\ref{StripSame}c) and another one metastable (Fig.~\ref{StripSame}d). The metastable branch bifurcates at $\epsilon \approx 0.055$ and terminates at $\epsilon \approx 0.025$, as seen most clearly in Fig.~\ref{StripSame}b. A branch of unstable shapes must exist within this interval but it cannot be located by our energy minimisation procedure.
	   	
	\begin{figure}[t]
		\begin{tabular}{cc} 
			(a)&(b)\\
			\includegraphics[width=.23\textwidth]{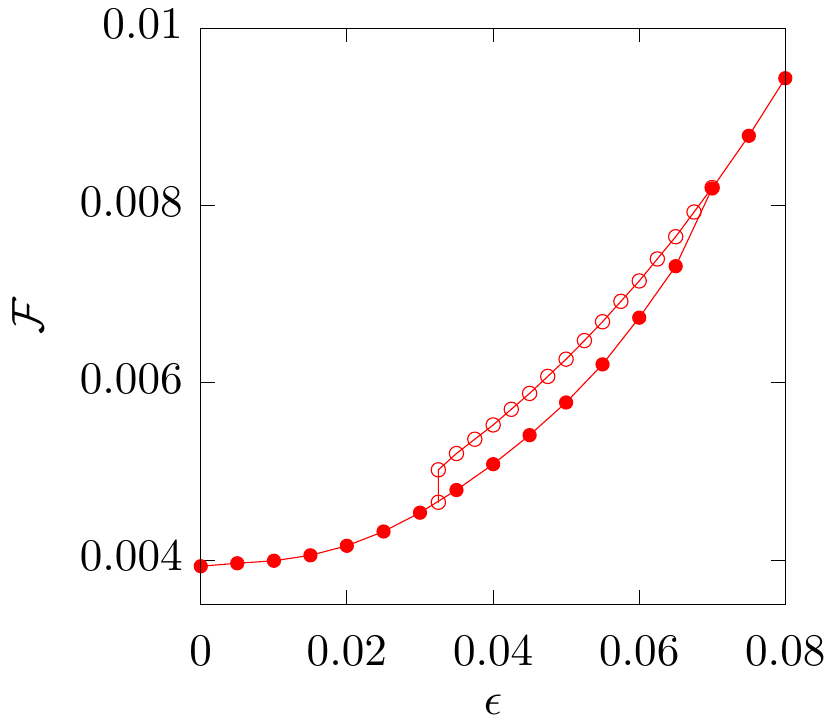}
			&\includegraphics[width=.22\textwidth]{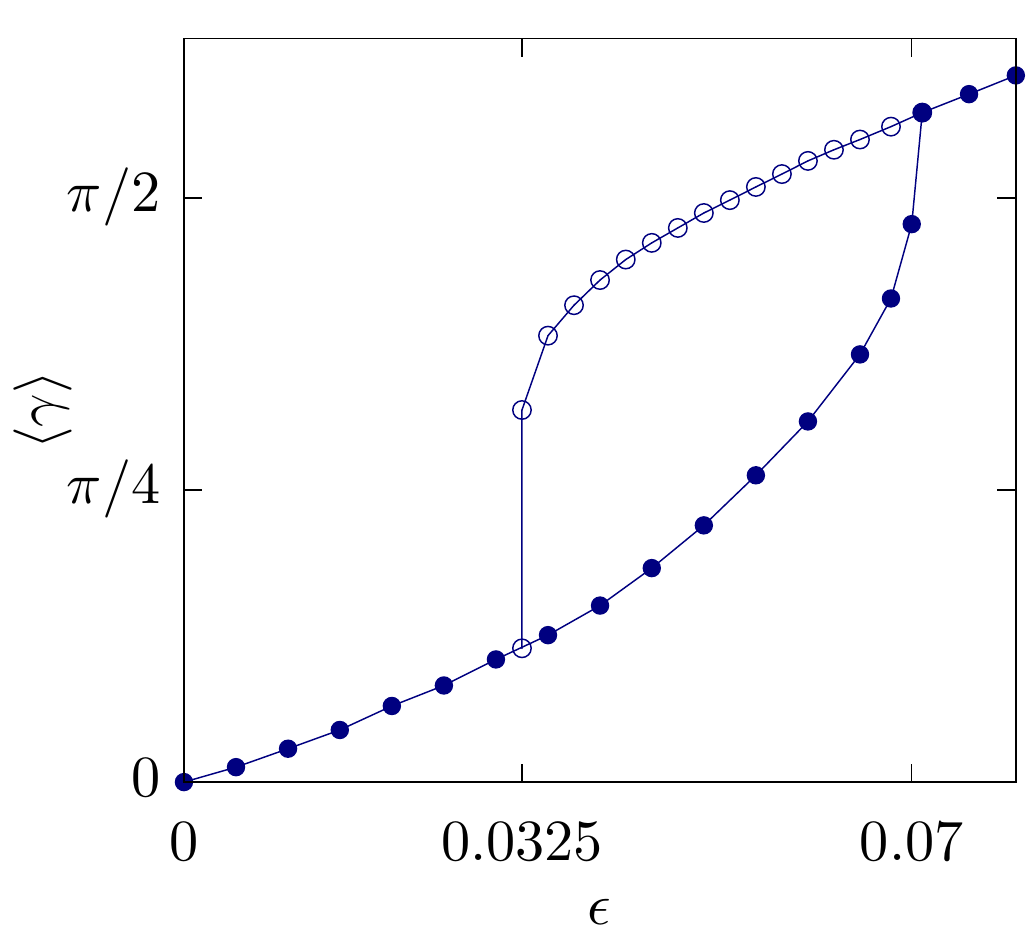}\\
			(c)&(d)\\
			\includegraphics[width=.225\textwidth]{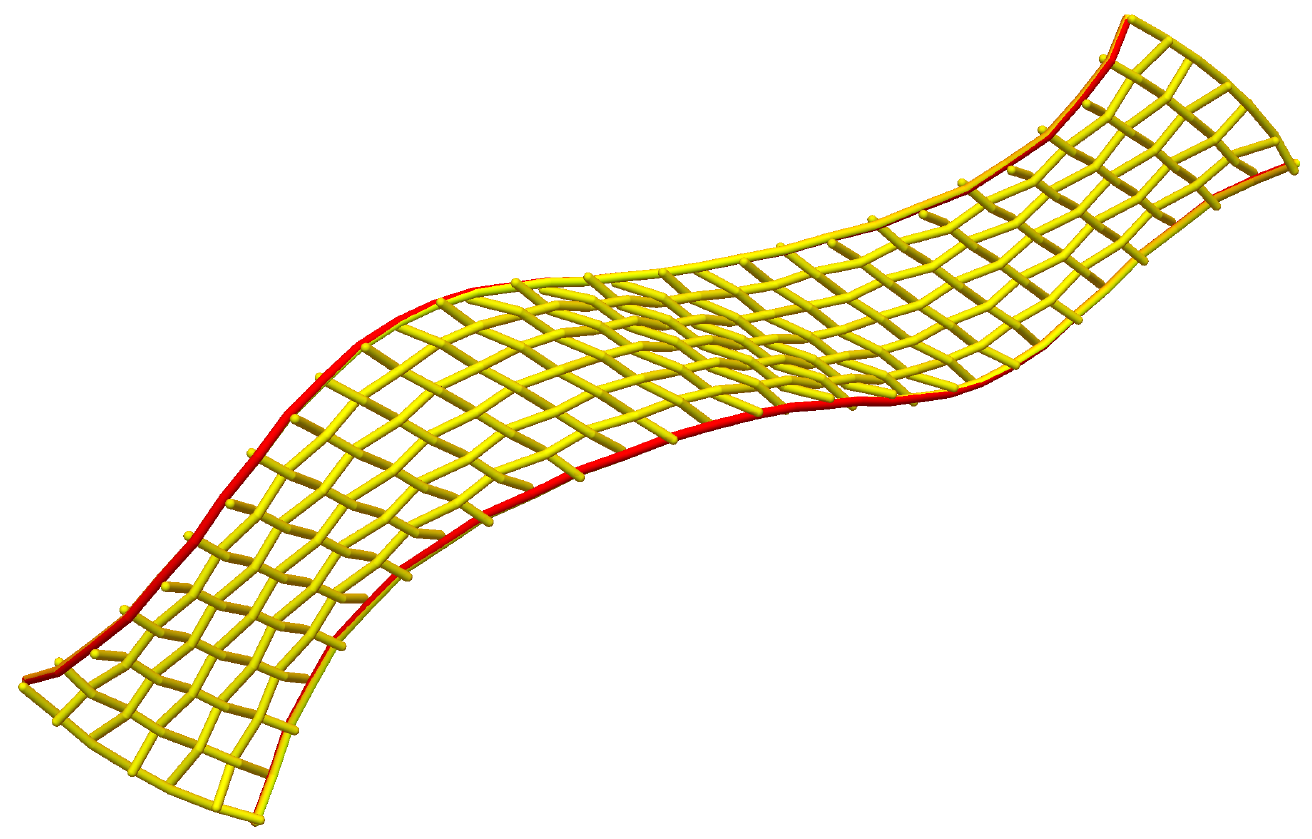}
			&\includegraphics[width=.225\textwidth]{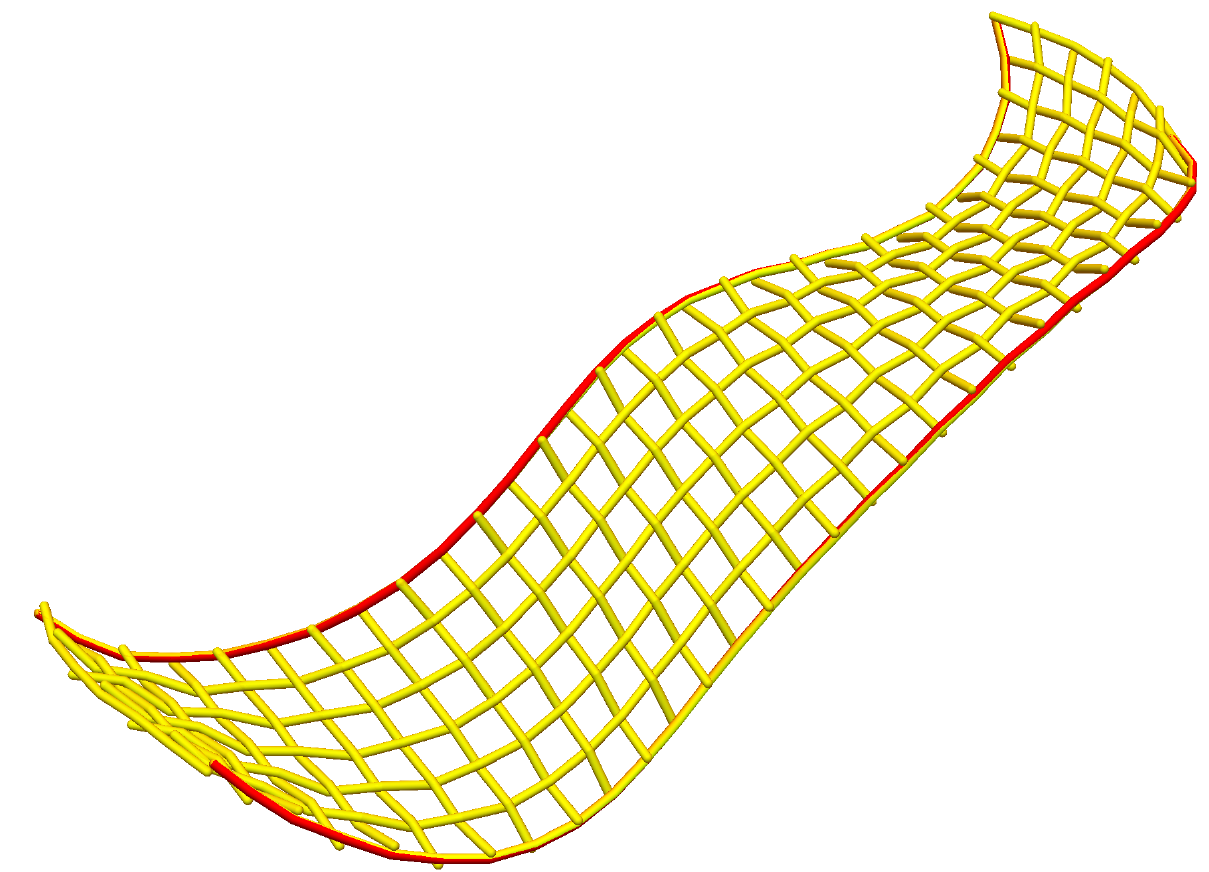}
		\end{tabular}
		\caption{The overall energy (a) and the average deviation angle $\langle \gamma \rangle$ (b) of two active fibres with different coiling handedness and the total rotation $2\pi$.  The equilibrium shapes of the stable (c) and metastable (d) configurations are shown at $\epsilon=0.04$.}
		\label{StripOpp}
	\end{figure} 

Another arrangement with multiple solutions involves active fibres with different coiling handedness, even with the same pitch length. As in the former case, the Janus filaments have misaligned intrinsic curvature vectors, so that the distances between them vary along the strip length when they bend upon activation, even though the average distance remains constant. The cause of multiplicity, as in the preceding case, lies in constraints imposed by the connecting fabric. At low extensions, the Janus fibres have opposite orientations at edges; in this case the projections of the curvature vectors on the original textile plane coincide and the shape is symmetric and unique. At high extensions, the solutions are unique as well but asymmetric, with a large angle $\langle \gamma \rangle$. One of active fibres turns with decreasing $\epsilon$ to reduce both $\langle \gamma \rangle$ and the overall energy (Fig.~\ref{StripOpp}c) at the critical value $\epsilon \approx 0.07$ (Fig.~\ref{StripOpp}a,b). The textile may remain in the less symmetric configuration (Fig.~\ref{StripOpp}d) even when the extension coefficient decreases below the critical value, so that the metastable branch forking off  there persists in the interval $0.0325<\epsilon<0.07$ terminating at the lower end of this interval where the symmetric state is restored. 

\section{Restricted twisted Janus rings \label{ring} }

Configurations with active fibres forming closed loops provide neat examples of stabilisation of closed configurations by the textile fabric and at the same time generate a variety of elaborate shapes. Isolated twisted closed fibres are absolutely unstable to off-plane deformations and acquire multiple convoluted shapes \cite{Jring}, but they are stabilised when restricted by passive fibres or woven into a fabric. Unlike a textile strip, closed active rings should have an integer pitch and may develop torsion. 

\begin{figure}[t]
 		\begin{tabular}{cc} 
 			(a)&(b)\\
 			\includegraphics[width=.225\textwidth]{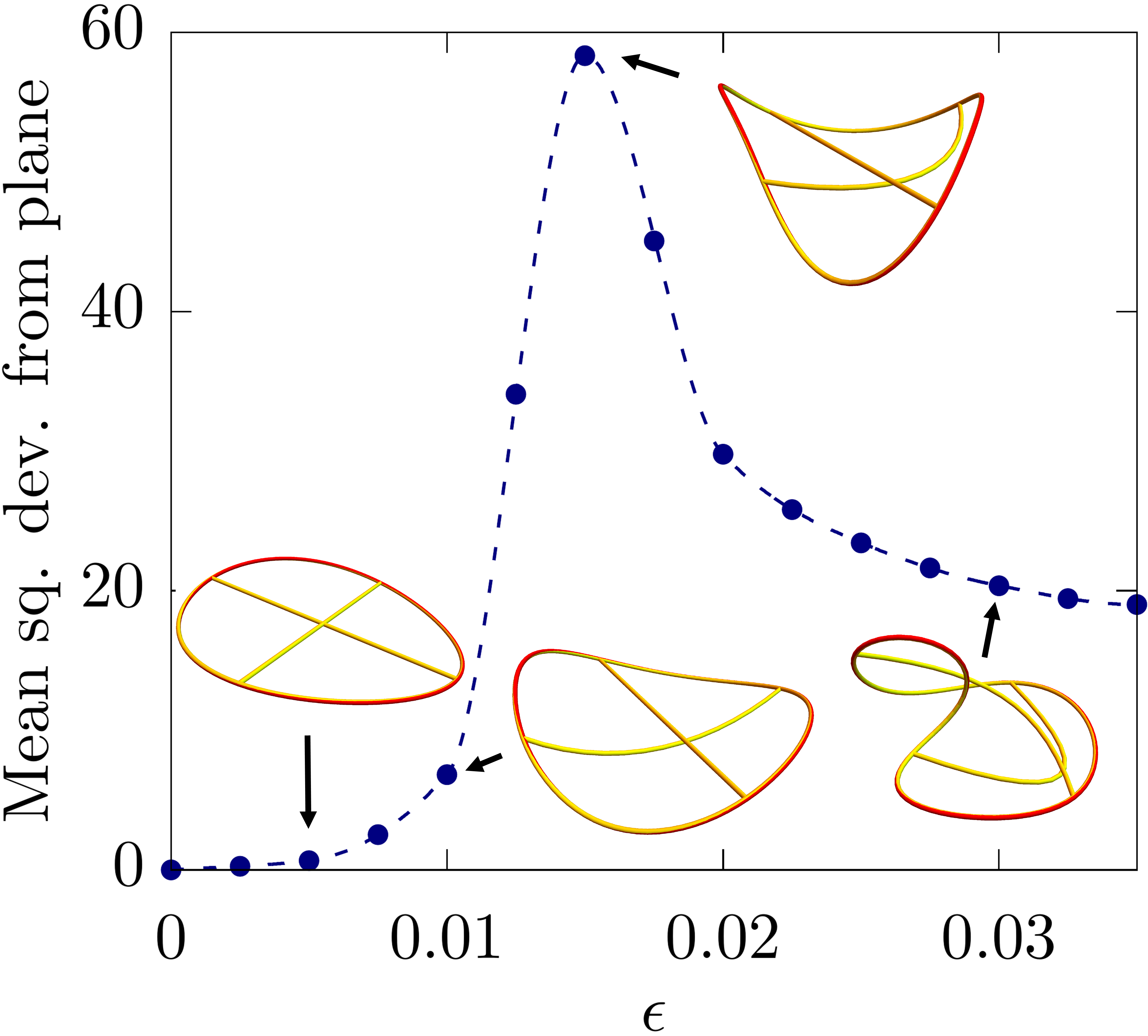}
 			&\includegraphics[width=.225\textwidth]{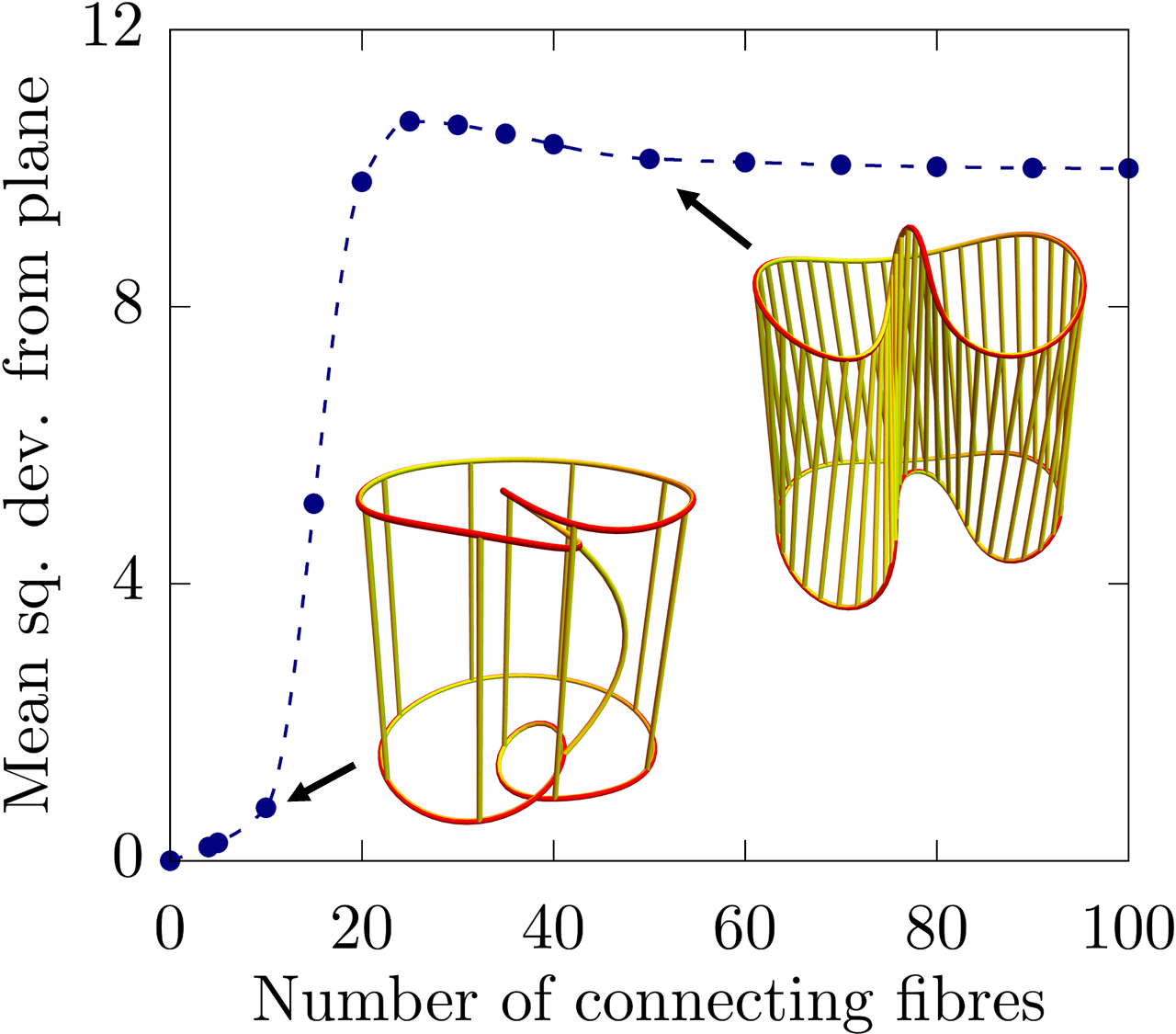}
 		\end{tabular}
 		\caption{(a) Reshaping of a twisted Janus ring restricted by two perpendicular passive fibres. (b) Deformation of two twisted Janus rings with opposite coiling handedness connected by a different number of passive fibres. }
 		\label{Restriction}
 	\end{figure}

A twisted Janus ring with two passive fibres with the lengths equal to the ring diameter crossing in the centre, which is initially in a planar configuration, passes through at least three qualitatively different configurations as the extension coefficient increases. The dependence of the deviation of the Janus fibre from original plane on the parameter $\epsilon$ is shown in Fig.~\ref{Restriction}a. At $\epsilon<0.01$, the bending energy of the ring is weak, and the passive fibres remain almost straight with minor deviations of the ring from the plane. Further increase of $\epsilon$ leads to bending of the passive fibres. First, at $0.01<\epsilon<0.015$, only one of them buckles, while the other one remains straight, causing the ring to acquire a significantly non-planar shape. With further increase, the entire structure inflects, so that at $\epsilon>0.015$ there is no stable configuration with both passive fibres remaining straight. Finally, the Janus ring produces a loop, as it develops an internal twist to relieve the bending energy. The average deviation from original plane decreases but, unlike an unconstrained Janus ring \cite{Jring}, it cannot fold up to a planar configuration, and warps until being restricted by a self-intersection limiting further deformation. 

The interaction between two twisted Janus rings connected by passive fibres also restricts deformations, as shown in Fig.~\ref{Restriction}b. The Janus rings have here the same length and pitch but opposite coiling handedness. We have compared the structures at the same value of the extension coefficient $\epsilon=0.03$ with a changing number $n$ of connecting passive fibres. The value $\epsilon=0.03$ is special, as a single ring with the internal normal rotated by $2\pi$ folds then to a double coverage of the circle with half the circumference. Such a folding would cause even a single connecting fibre to strongly bend but the elastic energy of Janus rings prevails over the energy of passive filaments when their number is small, and at $n\leqslant 5$ the rings retain the form of almost flat partially folded loops. At larger $n$, the increasing connectivity prevents bending of passive filaments and leads to growing deviations of the rings from the planar configuration, forming  loops perpendicular to the original plane. The shapes at selected values of $n$ are shown in Fig.~\ref{Restriction}b together with the plot of the average squared deviation of the rings from the average vertical position.
 
\section{Woven sleeves \label{cylinder} }

\begin{figure}[t]
 		\begin{tabular}{ccc} 
 			(a)&(b)&(c)\\
 			\includegraphics[width=.15\textwidth]{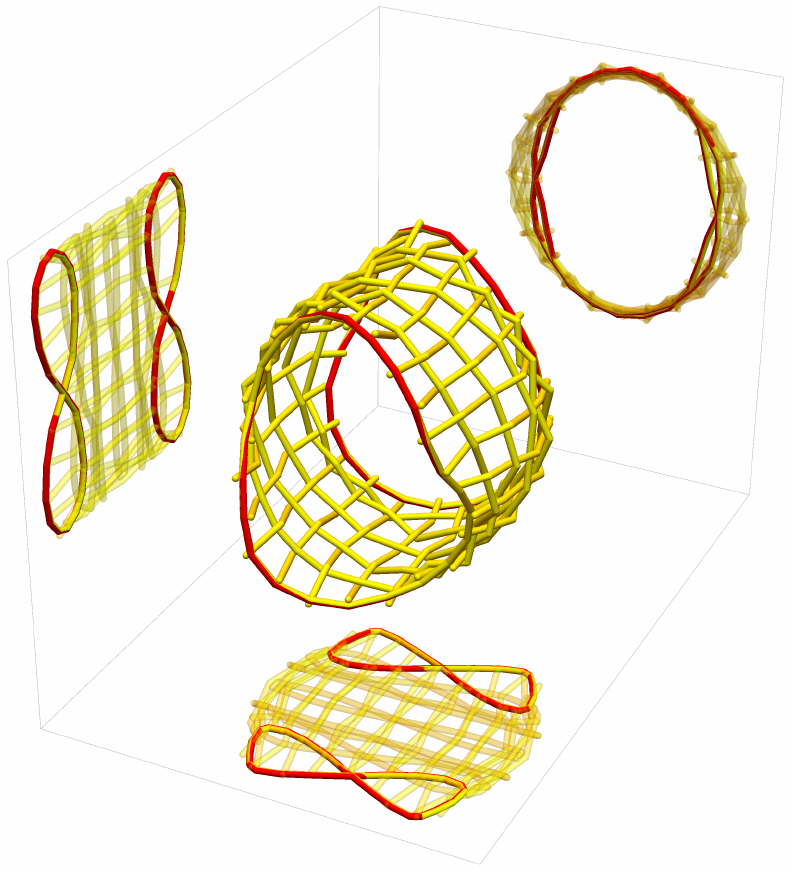}
 			&\includegraphics[width=.15\textwidth]{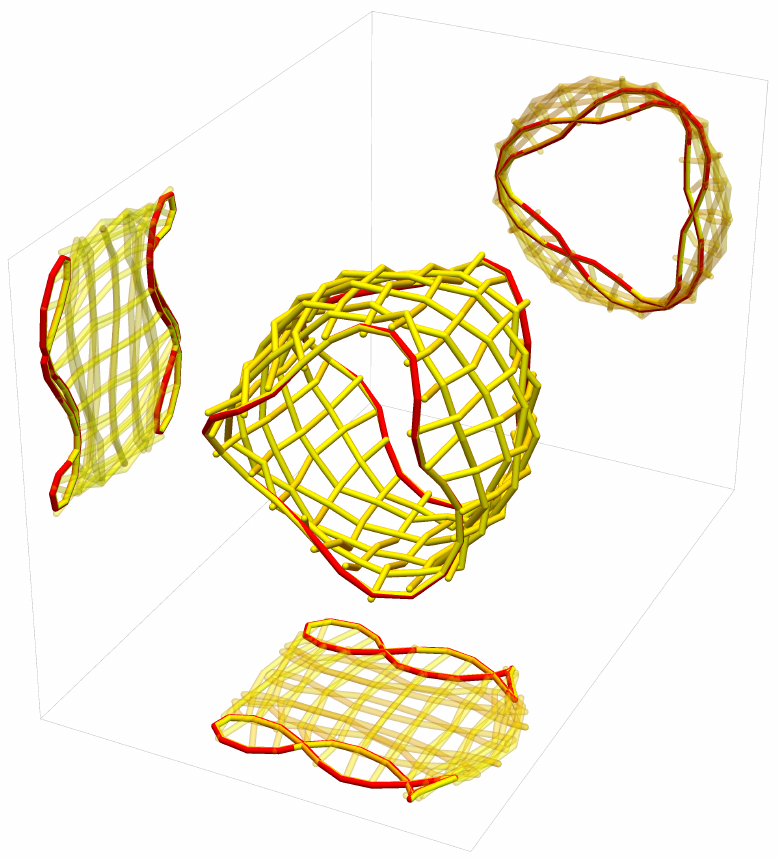}
 			&\includegraphics[width=.15\textwidth]{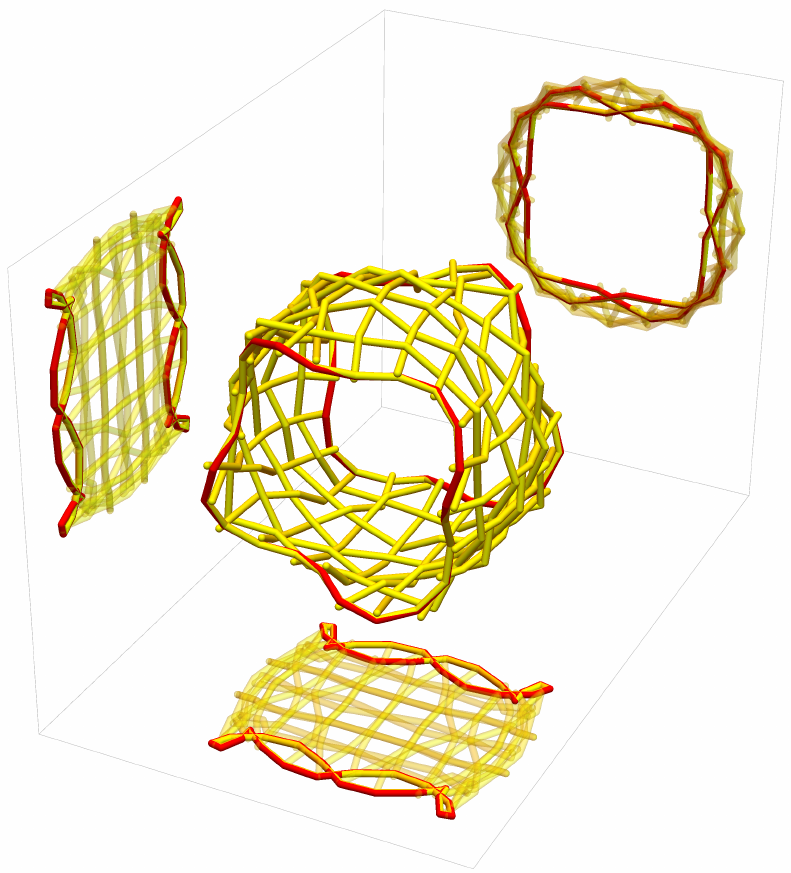}
 		\end{tabular}
 		\caption{Shapes of a cylindrical textile with two active fibres of the same handedness and (a-c) 2, 3, 4 turns of the internal normal, respectively, at the same extension parameter $\epsilon=0.06$. Projections on the three normal planes are also shown.}
 		\label{CylinderP}
 	\end{figure}

The constraints on the reshaping of Janus fibres become still more pronounced in woven cylindrical ``sleeve" structures. The computations with two identical active rings of the same handedness and pitch embroidered into a cylindrical textile are presented in Fig.~\ref{CylinderP}. The passive fibres tend to damp the deformations, and therefore the two active filaments, unlike unconstrained twisted Janus rings, are prevented from attaining an optimal curvature by producing multiple loops. This limitation causes the cylindrical fabric to deform into an ellipsoid, triangular or square cross-sectional shapes, depending on the number of full internal rotations of the Janus ring. 
    
\begin{figure}[t]
 		\begin{tabular}{cc} 
 			(a)&(b)\\
 			\includegraphics[width=.225\textwidth]{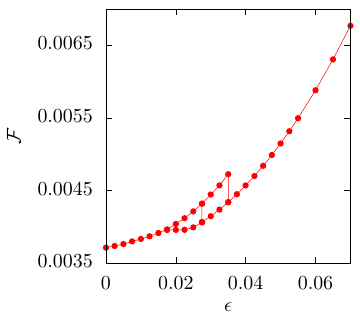}
 			&\includegraphics[width=.225\textwidth]{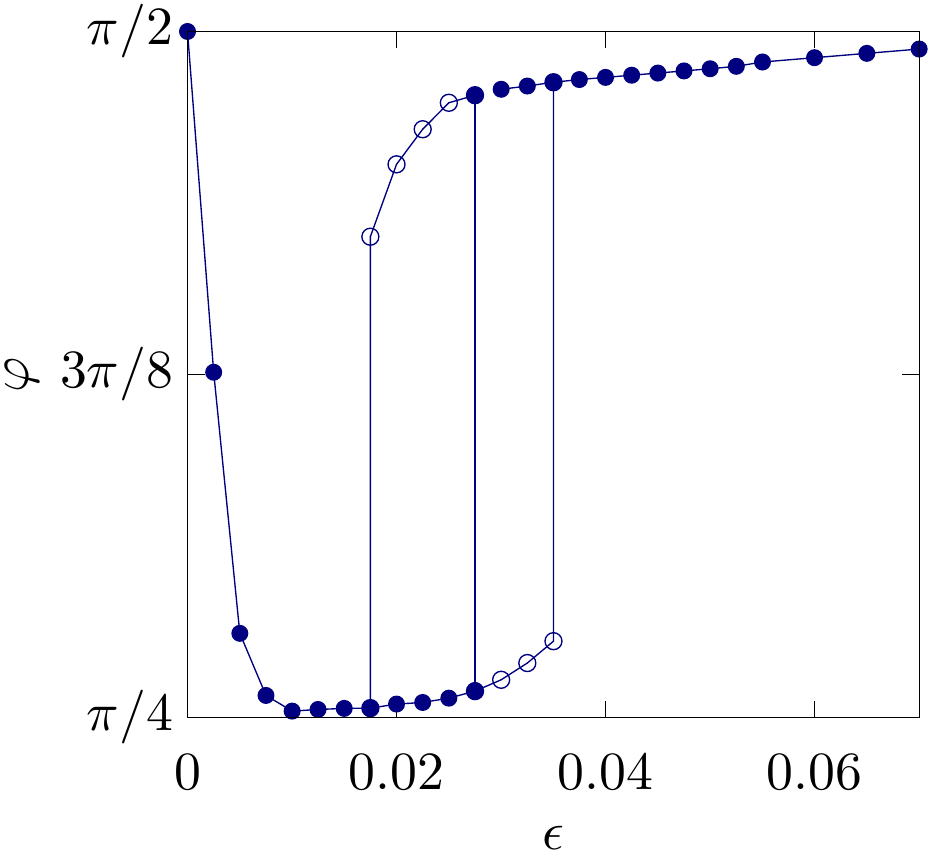}\\
 			(c)&(d)\\
 			\includegraphics[width=.16\textwidth]{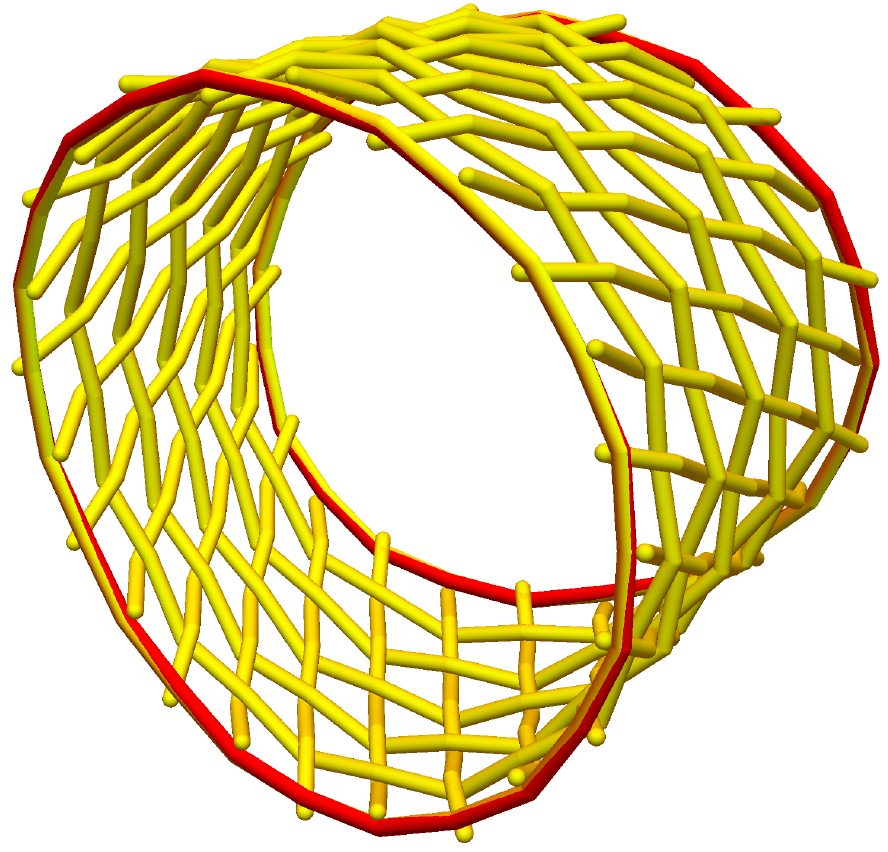}
 			&\includegraphics[width=.15\textwidth]{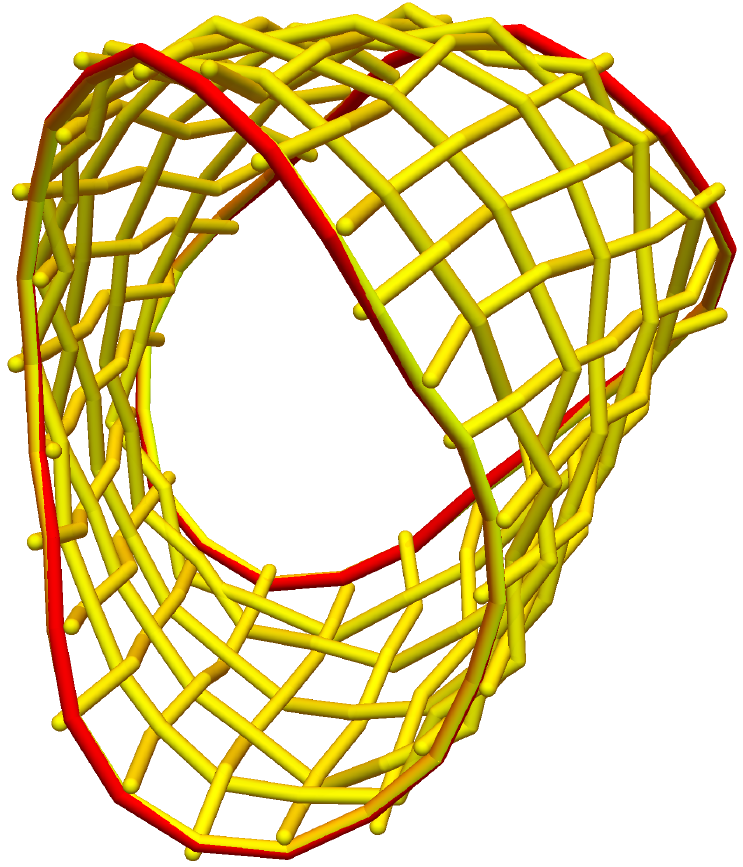}
 		\end{tabular}
 		\caption{The overall energy (a) and the average skew angle $\varphi$ (b) of a cylindrical textile of the radius $R=4$ framed by two Janus rings of different coiling handedness and rotation $4\pi$.  The respective shapes are shown at $\epsilon=0.015$ (c) and $\epsilon=0.05$ (d).}
 		\label{CylinderEps}
 	\end{figure}

 	Different handedness of the active fibres, similar to the arrangement discussed above for a strip, leads to the incompatibility of the distances between two initially parallel active rings. Fig.~\ref{CylinderEps} shows the results of computations for a cylindrical textile framed by two active rings rotated by $4\pi$ with opposite handedness. Two branches of solutions with different energy exist in the interval $0.0175\lesssim \epsilon \lesssim 0.035$. We characterise the shape by the average angle $\varphi$ between the filaments crossing at the textile nodes. The skewed structure with a large $\varphi$ is metastable but it persists in the multistability interval and is attained when the evolution starts from appropriate initial conditions.
 	
 \begin{figure}[t]
  		\begin{tabular}{cc} 
  			(a)&(b)\\
  			\includegraphics[width=.225\textwidth]{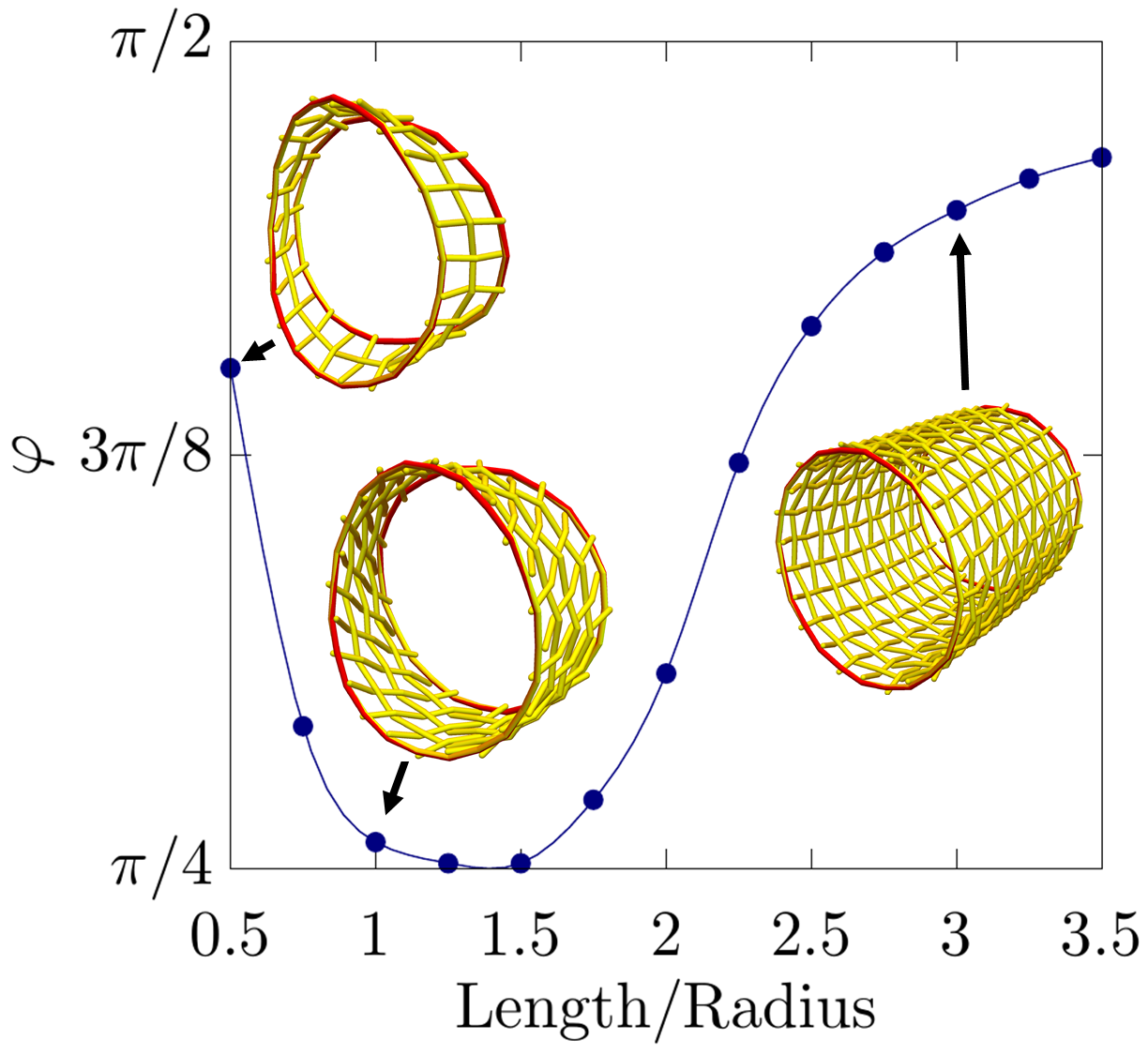}
  			&\includegraphics[width=.225\textwidth]{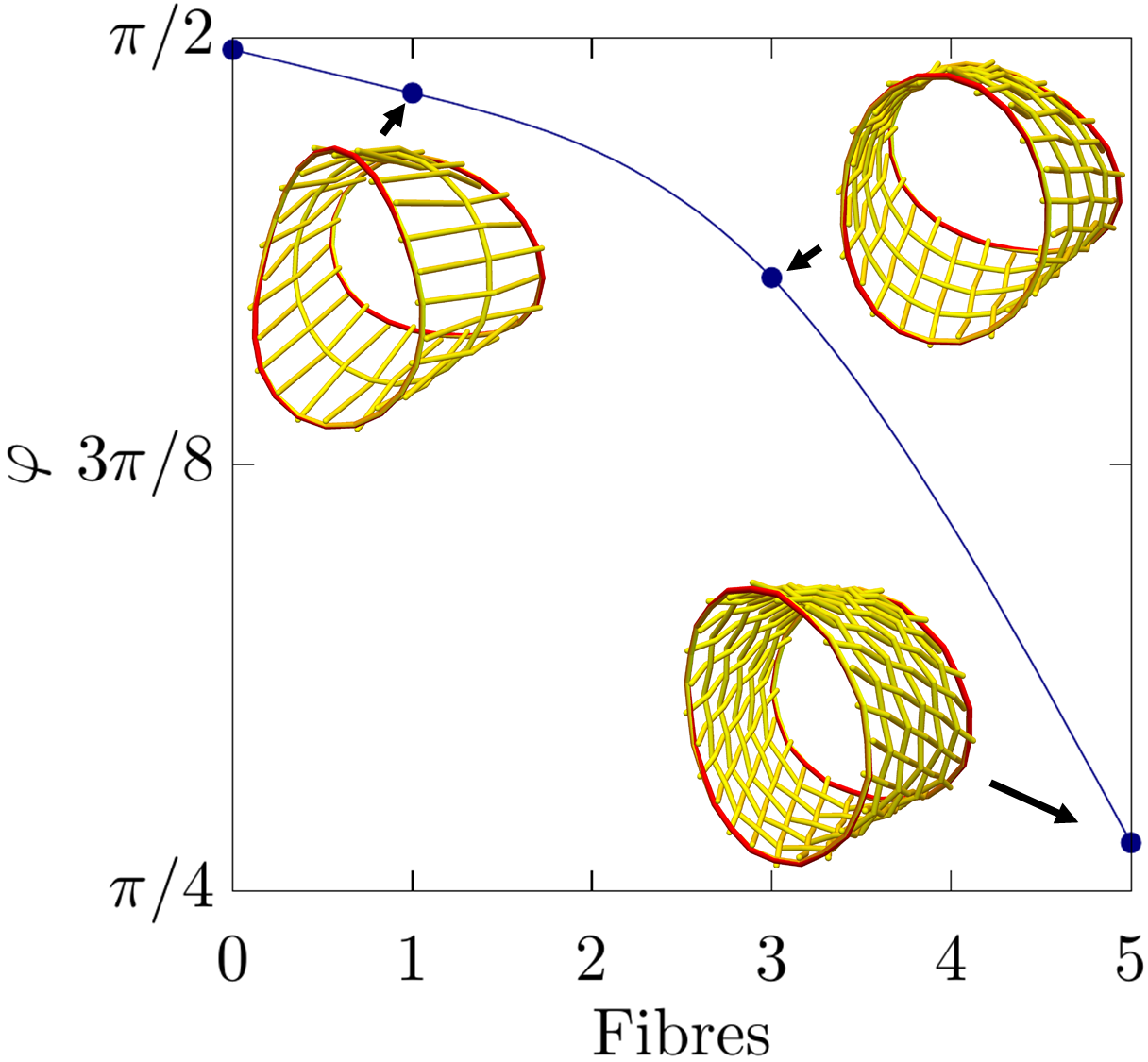}
  		\end{tabular}
  		\caption{The dependence of the angle $\varphi$ on the aspect ratio of the cylinder $L/R$ at $R=4$, $\epsilon=0.01$ (a) and on the number of parallel passive fibres at $L=6$, $R=4$, $\epsilon=0.03$ (b).}
  		\label{CylinderL}
  	\end{figure} 
  	
 We found that the angle $\varphi$ depends on the ratio between the length $L$ and the radius $R$ of the sleeve (measured in the units of length between textile nodes), approaching a maximum when the length is comparable to the radius (Fig.~\ref{CylinderL}a). In a squat cylinder, the textile can easily bend by changing its radius, so that it warps similar to a closed narrow ribbon, causing passive parallel fibres to bend and the perpendicular ones to skew. Bending of passive parallel fibres becomes too energetically expensive when the sleeve length increases to about twice the radius, so that the active fibres mostly buckle in-plane while the perpendicular passive filaments tend to be parallel to the cylinder axis ($\varphi \to \pi/2$). These changes are continuous and no multistability is observed.
 	
Another shape transition takes place when the textile is subjected to a connectivity change. We have run simulations starting from the metastable state at $\epsilon=0.03$ in Fig.~\ref{CylinderEps} and gradually reducing the number of parallel passive fibres which stabilise the shape. As thinning the textile reduces the interconnection strength, the structure does not remain in a metastable state. We found that the transition to the stable state is not influenced by the number of perpendicular connecting fibres, but is strongly affected by reducing the number of parallel passive fibres (Fig.~\ref{CylinderL}b). This effect is similar to the dependence on the cylinder length: the active rings can bend significantly normally to the textile plane as the resistance by the parallel passive fibres weakens, and the textile finally relaxes to the conformation with  straight perpendicular filaments ($\varphi=\pi/2$).

\section{Conclusion}

Adding internal twist to Janus filaments, which causes direction of bending to vary along their length, leads to a great variety of shapes attainable upon actuation. It also enhances multistability of shapes, which we detected already in simple configurations studied here. On the other hand, the fabric of passive filaments stabilises internally twisted fibres preventing highly convoluted shapes. Structures of this kind may prove useful whenever controlled reshaping is desirable.

\section*{Acknowledgements}

 This research is supported by Israel Science Foundation (grant No. 669/14). A.P.Z. also acknowledges the support by RFBR according to research project No. 18-31-00023.


\begin{thebibliography}{99}

\bibitem{textile} A.~P. Zakharov and L.~M. Pismen, Soft Matter \textbf{14}, 676 (2018).

\bibitem{GorielyTabor} A. Goriely and M. Tabor, Nonlinear Dyn. \textbf{21}, 101-133 (2000).

\bibitem{GorielyGoldstein} R. E. Goldstein and A. Goriely, Phys. Rev. E \textbf{74}, 010901 (2006).

\bibitem{Liu15}J. Liu, J. Huang, T. Su, K. Bertoldi, and D. Clarke, PLoS ONE 9 (4), e93183 (2014).

\bibitem{Audoly17} C. Lestringant and B. Audoly,  J. Mech. Phys. Solids \textbf{103} 40-71 (2017) .

\bibitem{Goriely17} T. Lessinnes, D. E. Moulton, and A. Goriely,  J. Mech. Phys. Solids \textbf{100} 147 (2017).

\bibitem{Goriely06}A. Goriely, J. Elasticity \textbf{84}: 281-299  (2006).

\bibitem{Michell} J. H. Michell, Messenger of Math. \textbf{19}, 68-82 (1889-90).

\bibitem{Jring} A.~P. Zakharov and L.~M. Pismen, Phys. Rev. E \textbf{97}, 062705 (2018).

\bibitem{IonovJanus} L. Ionov, G. Stoychev, D. Jehnichen, and J. U. Sommer, Appl. Mater. Interfaces \textbf{9}, 4873 (2017).

\bibitem{LL} L. D. Landau and E. M. Lifshitz, \emph{Theory of Elasticity} (Pergamon Press, Oxford, 1970).

\bibitem{AP}
B.~Audoly and Y.~Pomeau, \emph{Elasticity and geometry: from hair curls to the non-linear response of shells} (Oxford University Press, Oxford, 2010).

\end{thebibliography}
\end{document}